\documentclass[sigconf]{acmart}
\AtBeginDocument{%
  }

\usepackage{enumitem}

\acmConference[Tools for Thought @ CHI]{}{April 26, 2025}{Yokohama, Japan}

\acmDOI{}
\acmISBN{}

\begin{document}

\title[Supporting AI-Augmented Meta-Decision Making with InDecision]{Supporting AI-Augmented\\Meta-Decision Making with InDecision}

\author{Chance Castañeda}
\authornote{Co-first authors contributed equally to this research.}
\affiliation{%
  \institution{Carnegie Mellon University}
  \city{Pittsburgh}
  \state{PA}
  \country{USA}
}
\email{chancec@andrew.cmu.edu}

\author{Jessica Mindel}
\authornotemark[1]
\affiliation{%
  \institution{Carnegie Mellon University}
  \city{Pittsburgh}
  \state{PA}
  \country{USA}
}
\email{jrmindel@andrew.cmu.edu}
\orcid{0000-0001-5506-216X}

\author{Will Page}
\authornotemark[1]
\affiliation{%
  \institution{Carnegie Mellon University}
  \city{Pittsburgh}
  \state{PA}
  \country{USA}
}
\email{wpage@andrew.cmu.edu}
\orcid{0009-0003-2007-6918}

\author{Hayden Stec}
\affiliation{%
  \institution{Carnegie Mellon University}
  \city{Pittsburgh}
  \state{PA}
  \country{USA}
}
\email{hstec@andrew.cmu.edu}

\author{Manqing Yu}
\affiliation{%
  \institution{Carnegie Mellon University}
  \city{Pittsburgh}
  \state{PA}
  \country{USA}
}
\email{manqingy@andrew.cmu.edu}

\author{Kenneth Holstein}
\affiliation{%
  \institution{Carnegie Mellon University}
  \city{Pittsburgh}
  \state{PA}
  \country{USA}
}
\email{kjholste@andrew.cmu.edu}
\orcid{0000-0001-6730-922X}

\renewcommand{\shortauthors}{Castañeda, Mindel, Page, et al.}

\begin{abstract}
From school admissions to hiring and investment decisions, the first step behind many high-stakes decision-making processes is ``deciding how to decide.'' Formulating effective criteria to guide decision-making requires an iterative process of exploration, reflection, and discovery. Yet, this process remains under-supported in practice. In this short paper, we outline an opportunity space for AI-driven tools that augment human \textit{meta-decision making}. We draw upon prior literature to propose a set of design goals for future AI tools aimed at supporting human meta-decision making. We then illustrate these ideas through \textit{InDecision}, a mixed-initiative tool designed to support the iterative development of decision criteria. Based on initial findings from designing and piloting \textit{InDecision} with users, we discuss future directions for AI-augmented meta-decision making.
\end{abstract}



\keywords{Meta-Decision Making, Decision Criteria, Reflection, Elicitation, Prototyping, Iteration, Augmentation, AI, LLM, Mixed Initiative}

\received{20 February 2025}
\received[accepted]{14 March 2025}


\maketitle

\section{Introduction}

In high-stakes decision-making contexts, such as school admissions, hiring, investment, and peer review, it is often valuable to develop \textit{explicit criteria} to guide decision-making~\cite{bridgespan2023developing,culpepper2023rubrics,donabedian1981advantages,greco2016multiple,kahneman2016noise}. Research shows that explicating decision criteria can lead to more systematic and consistent decisions both in individuals (by encouraging the use of the same criteria when evaluating different options) and across groups (by encouraging use of consistent criteria across different decision-makers)~\cite{donabedian1981advantages,greco2016multiple,kahneman2016noise}. Explicating criteria can also increase transparency, enabling inspection and critique in contexts that require it, which may help to promote cooperation among decision-makers.

The flip side of these benefits, however, is that the use of poor quality decision criteria risks producing systematically poor decisions~\cite{culpepper2023rubrics,donabedian1981advantages,mcmackin2000does,zhang2023deliberating}. In practice, it is often challenging to develop decision criteria that capture what truly matters to an individual or group, for several reasons that we outline here. In the context of complex decision tasks, people typically have limited conscious or explicit knowledge about what they actually value~\cite{newell2014unconscious}. As a result, when directly asked to come up with decision criteria, the criteria people propose initially tend to be only dubiously connected to what they truly care about in a decision. Often, people only ``know it when they see it''—they learn about their own criteria only when directly comparing different concrete options. As people compare more options, it is common for criteria to evolve~\cite{kuo2024policycraft,lam2024ai,pirolli2005sensemaking,liu2023selenite,shankar2024validates}, a phenomenon Shankar et al. have called \textit{criteria drift} \cite{shankar2024validates}. As a concrete example: in the context of university faculty hiring or graduate admissions, criteria are often developed through abstract discussions about future goals and priorities (e.g., \textit{``What do we want our department to be known for in five years?'', ``Where do we want to grow our expertise?'', ``What kinds of students do we want in our program?''}). However, as committee members review more and more candidates, it is common to notice gaps between their stated decision criteria versus what actually excites them about individual candidates.

Thus, effectively formulating explicit criteria requires an iterative process of reflection and discovery, akin to the reflective processes integral to prototyping~\cite{schon1992designing}. This involves continuously assessing how well abstract criteria align with one's actual evaluations of specific, concrete instances~\cite{pirolli2005sensemaking,shankar2024validates}. Today, this reflective process is not typically well supported in practice. Decision criteria are often developed in just one or a few iterations, and are evaluated in the abstract rather than being prototyped and stress-tested against realistic options. Given time pressure to move ahead to decision-making, early drafts of criteria become prematurely entrenched as official process. Decision-makers have limited opportunities to test how well their decision criteria work and to iterate on criteria before using them.

\textbf{We see a rich opportunity space for AI-driven tools to support the human process of ``deciding how to decide,''} known in psychology as \textit{meta-decision making}~\cite{boureau2015deciding,mintzberg1976structure,remus1986toward}. Past HCI research has explored AI tools that augment human decision-making, such as tools offering in-the-moment recommendations, nudges, or sensemaking support to aid specific decisions (e.g.,~\cite{buccinca2021trust,kawakami2022improving,lai2023towards,liu2023selenite,ma2024towards,yang2019unremarkable}). However, comparatively little work has explored the use of AI to support individuals and groups in exploring, reflecting on, and articulating what truly matters to them in future decision-making~\cite{kuo2024policycraft,zhang2023deliberating}. In the remainder of this paper, we first propose a set of design goals for future AI tools aimed at supporting human meta-decision making, drawing on relevant concepts and theories across disciplines. To illustrate these ideas more concretely, we then introduce \textit{InDecision}, a mixed-initiative tool designed to support the iterative process of developing decision criteria. Informed by initial findings from designing and piloting \textit{InDecision} with users, we discuss future directions for HCI research on AI-augmented meta-decision making.

\section{Toward AI-Augmented Meta-Decision Making}

Building on prior literature in HCI, psychology, learning sciences, and decision science, we propose a set of high-level design goals for AI tools that support human meta-decision making: 

\begin{enumerate}[label=D\arabic*.,start=1]
    \item \textbf{Position humans as judges and AI as provocateur.} Viewed as a reflective process, where the goal is to help people learn what \textit{they} care about in future decisions, augmenting human meta-decision making necessarily relies on judgments and knowledge that are internal to users (e.g., \textit{``this isn't the sort of candidate we'd usually consider... but it feels like admitting them would create a really interesting opportunity for the program''}). As such, tools must be designed to treat humans as the ultimate judges when it comes to evaluating decision criteria or concrete options. AI tools can help this process by stimulating reflective thought and discussion~\cite{cai2024antagonistic, drosos2025makes, liu2023selenite}, provoking users to reconsider how they have externalized their decision criteria so far. For example, AI-based provocations might aim to induce curiosity, encouraging users to explore a certain line of questioning further, or they might aim to induce dissonance by surfacing potential incongruencies between externalized decision criteria and users' actual judgments on concrete options~\cite{harmon-jones2019dissonance, mcgrath2017dealingdissonance}. To do so, designers might create feedback loops in which users' judgments are used as data to inform generated provocations. \\
    
    \item \textbf{Engage users in making decisions on concrete options, to support reflection on decision criteria.} AI tools for meta-decision making should present tangible examples of decision options---whether real or generated---to help users understand abstract dimensions of a decision. These examples offer reference points that can be readily evaluated, accepted, or challenged~\cite{compton2015casual}, helping users ``know it when they see it''. AI tools for meta-decision making should encourage users to assess and compare concrete options early and often, to help them reflect on the alignment between their decisions and how they have externalized their criteria so far~\cite{kuo2024policycraft,shankar2024validates}.
     
\end{enumerate}
To ensure that the concrete examples a tool provides are helpful in prompting reflection and iteration, tools should:

\begin{enumerate}[label=D\arabic*.,start=3]
    \item \textbf{Systematically vary presented options to help users discover new dimensions.} Variation Theory~\cite{marton2014necessary}, a theory of human concept learning, suggests that comparison among diverse instances can help people identify meaningful dimensions of similarity and difference (i.e., axes of variation). In the context of augmenting human meta-decision making, AI tools should leverage users' past assessments of both options and criteria to \textit{systematically vary} presented options, with the goal of helping users discover and reflect upon potential dimensions~\cite{suh2024luminate} that matter to them. \\
    
    \item \textbf{Generate potential ``edge cases'' to stress-test criteria.} Tools should generate edge cases~\cite{drosos2025makes,kuo2024policycraft,zhang2023deliberating} likely to reveal differences between a user's current, stated decision criteria and the way they actually evaluate concrete cases. For example, a tool might generate a case that sits at a (so far) unexplored intersection of multiple potentially competing decision criteria, in order to probe further on the relationship between these criteria. \\
    
    \item \textbf{Enable iteration on all elements of the decision, including how the decision itself is conceptualized.} Throughout the iterative process of developing decision criteria, all elements of a decision can evolve, including not only the criteria but also the options under consideration and even how the decision is framed. Tools for meta-decision making should enable users to flexibly iterate on each of these elements throughout the process---for example, through interactive interfaces that hierarchically organize~\cite{suh2023sensecape} relevant criteria and options within alternative framings of the decision at hand, and support users in navigating between these alternative framings as needed.
\end{enumerate}

Finally, to help users communicate about their reflection:

\begin{enumerate}[label=D\arabic*.,start=6]
    \item \textbf{Preserve and support reflection on process.} Making process visible helps people reuse earlier work to inspire new directions, accurately reflect on their thought process, and gain confidence in their progress~\cite{kery2018story, sterman2022version}. Rather than solely presenting final drafts of criteria, tools should make changes in elements transparent, navigable, shareable, and revisitable (e.g., branching from a version), ensuring that it is possible for users to quickly trace back the considerations and inspirations that led to these changes~\cite{kery2018story}. Tools may also present summaries or overviews of the criteria development process, to scaffold reflection and conversation about process. \\
    
    \item \textbf{Mitigate social pressure to conform.} To prevent time pressure or unequal power dynamics from leading to premature convergence, tools should ensure that individuals  have the opportunity to think about and share their perspective to encourage \textit{productive dissensus}~\cite{chen2023judgment,koshy2024venire,kuo2024wikibench}. For example, in group contexts, tools might play a mediating role by helping to orchestrate discussions or by presenting collaborators' ideas on their behalf.
\end{enumerate}

\section{InDecision: Helping People Develop Decision Criteria through Iterative Prototyping}

\begin{figure*}
    \centering
    \includegraphics[width=0.9\linewidth]{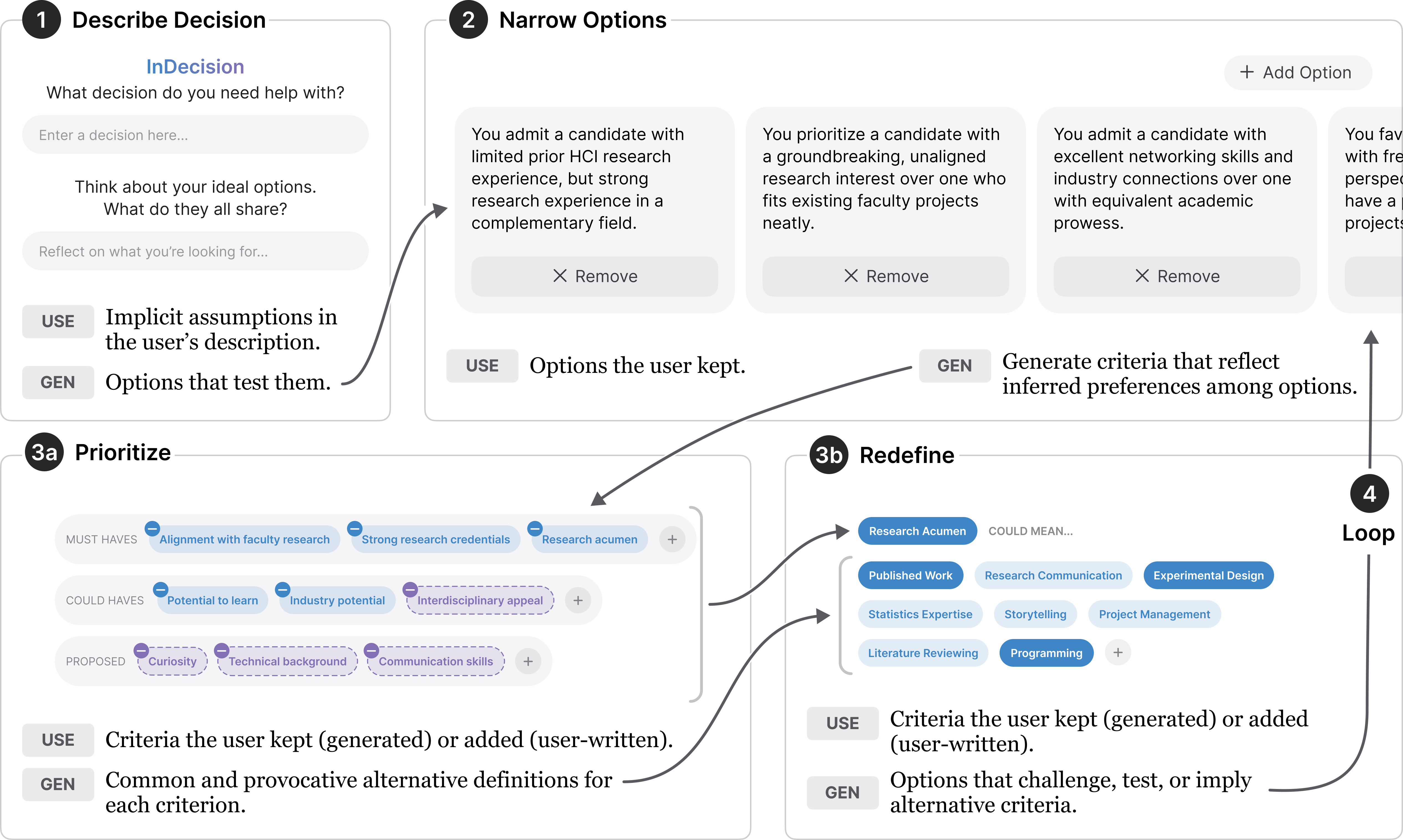}
    \caption{\textit{InDecision's} iterative loop. The initial elicitation (1) allows the user to provide open-text descriptions of their decision and relevant considerations for options and criteria. The user is presented with a list of eight options (2). The user can keep, add to, or remove these options. To promote reflection on what is most important to them, the user may continue only after narrowing down to three. Criteria refinement is composed of two stages: (3a) prioritization, where a user can add, remove, and sort criteria in tiers of priority; and (3b) redefinition, where the user selects between a range of possible meanings associated with each criterion. These steps are repeated in an iterative loop (4).}
    \label{fig:indecision}
    \Description{Figure 1 shows an overview of InDecision's iterative loop to show what data the user and large language model are engaging with in each step of the loop. It lists four steps. Step 1 is titled Describe Decision and includes two questions with textboxes for the user: ``What decision do you need help with?'' and ``Think about your ideal options. What do they all share?'' The system uses implicit assumptions in the user's responses in order to generate concrete options that test them. It displays these options in step 2, titled Narrow Options, as a set of cards with text and a button that toggles between remove and keep, as well as an option for the user to write their own option cards. The example options shown in the image are as follows. First, ``You admit a candidate with limited prior HCI research experience, but strong research experience in a complementary field.'' Second, ``You prioritize a candidate with a groundbreaking, unaligned research interest over one who fits existing faculty projects neatly.'' Third, ``You admit a candidate with excellent networking skills and industry connections over one with equivalent academic prowess.'' Based on which options the user chose to keep, the system generates criteria that reflect inferred user preferences among options. It displays these in step 3a, titled Prioritize, in which users can drag and drop proposed criteria, or ones that they write themselves, into one of three bins: must-haves, should-haves, and could-haves. The example criteria shown in the figure are as follows: alignment with faculty research, strong research credentials, research acumen, potential to learn, industry potential, interdisciplinary appeal, curiosity, technical background, and communication skills. Based on which criteria the user chose to keep, the system generates both common and provocative alternative definitions for each criterion. It displays these alternative definitions for each criterion from step 3a in step 3b, titled Redefine. The example alternative definitions shown for ``research acumen'' in this figure are as follows: published work, research communication, experimental design, statistics expertise, storytelling, project management, literature reviewing, and programming. Users can select and write their own alternative definitions that they would like to use as criteria going forward. Based on the criteria that the user keeps, the system returns to step 2, Narrow Options, with a new set of generated options that challenge, test, or imply these alternative criteria.}
\end{figure*}

As a concrete case study, we introduce \textit{InDecision}, a mixed-initiative tool that helps users rapidly, iteratively prototype their decision criteria. As overviewed in Figure~\ref{fig:indecision}, \textit{InDecision} aims to help users iteratively develop a set of decision criteria by stress-testing them against a simulated decision space. Where relevant, we link aspects of \textit{InDecision}'s design to the goals outlined above.

\begin{enumerate}
    \item \textbf{Describe Decision}: The \textit{InDecision} interface first prompts users to briefly articulate the context of their decision, and what they think they are looking for in ideal options [\textbf{D1}]. For example, in the context of graduate admissions decisions, a user might briefly describe general qualities they are looking for in candidates. The information elicited from users at this step, and at subsequent steps of the workflow, is intended to both (1) support the generation of more relevant simulated decision options and suggested criteria, and (2) identify potential gaps or misalignments in users' externalized criteria [\textbf{D1, D2, D4}].
    \item \textbf{Narrow Options}:  The current version of \textit{InDecision} uses information gathered from this initial elicitation step to generate a diverse starter set of eight concrete decision options. These options are generated by an LLM with the aim of challenging the user to reconsider their initial stated decision criteria [\textbf{D1-D4}], for example by introducing potential ``edge cases'' [\textbf{D3}]. To support this reflection, users are asked to narrow these eight options to the three that most resonate with them [\textbf{D2}].
    \item \textbf{Refine Criteria}: Leveraging users' previously stated decision criteria and their selections of concrete options as input data, \textit{InDecision} next generates a set of up to six inferred criteria that might better explain their selections [\textbf{D1}]. Users refine this set of criteria through the following steps:
    \begin{enumerate}
        \item \textbf{Criteria Prioritization:} Users are asked to rate how important different criteria are to them. As part of this process, they also have the ability to remove criteria or add new criteria that they feel are missing [\textbf{D5}].
        \item \textbf{Criteria Redefinition:} For each criterion that users decide to keep in their set, the system generates a set of alternative definitions intended to prompt reflection and elaboration on what they truly mean by that criterion [\textbf{D1}]. For example, if a user includes ``research acumen'' as a criterion, the system might offer more specific variations, such as: ``published work,'' ``research communication,'' or ``experimental design.''
    \end{enumerate}
    \item \textbf{Iterate}: Steps 2 and 3 repeat in a loop, carrying over any criteria that users chose to keep during the refinement stage. The system generates 8 new option provocations that each align with or challenge these criteria, encouraging users to continue iterating. This loop continues until users converge on a set of criteria that they feel confident in, despite the system's continued attempts at provocation and stress-testing. 
\end{enumerate}

\subsection{Design Exploration and Initial Findings}
We iteratively co-designed and piloted \textit{InDecision} with an initial set of 11 participants. Throughout this process, we explored alternative ways of representing decision spaces and surfacing tacit criteria through provocation. 
In each pilot, participants were asked to use the tool to think through an important upcoming decision in their life. Below are some of our early insights from these pilot studies, which point to directions for future research and design:

\subsubsection{A need for fluid transitions between iteration on options and iteration on criteria.} As described above, in the current version of \textit{InDecision}, users first evaluate options (Step 2) and then refine criteria (Step 3) in an iterative loop. However, during think-alouds, we found that users naturally generated ideas for new or modified criteria while evaluating the available options in Step 2; and they reflected on additional relevant options when considering criteria in Step 3. By the time users moved on to the next step of the workflow, most of these reflections were forgotten. These observations suggest a need for greater fluidity: tools should support users in continuously iterating on \textit{both} their decision criteria and their understanding of the decision space, from any point in their workflow. For instance, rather than ordering Step 2 and 3 in a strict sequence, criteria and option provocations might be displayed side-by-side throughout the entire workflow. Such a design could have the added advantage of helping users more immediately see and reflect upon potential misalignments between their externalized decision criteria and their actual judgments on concrete options.

\subsubsection{Representing relationships across criteria.} While we began with the notion that users would iterate upon sets of \textit{independent} or separable criteria, users often thought about \textit{structures} of interrelationships when developing decision criteria. As such, participants sometimes found it limiting to explore and prioritize potential decision criteria through importance ratings or rankings alone. Instead, they gravitated toward interactive visualizations that afford representing non-additive interactions or dependencies, whether through clustering, parent-child relationships, or conditional preferences. This suggests an opportunity for future work on AI tools for meta-decision making to explore more expressive representations that can capture rich structure among criteria (cf.~\cite{liu2023selenite}).

\subsubsection{Expanding the space of options that decision-makers are able to consider.} While users were quick to reject options that seemed unrealistic to their situation, many were also surprised to find that the system expanded their understanding of the decision space. For example, in cases where participants had initially framed a decision as a binary choice, \textit{InDecision} offered additional concrete and relevant options that they had not previously considered. These options often sat somewhere in between the choices a participant described initially, blending advantages of multiple options.  
Although the current design of \textit{InDecision} is focused mainly on presenting option provocations to help people iterate on \textit{decision criteria}, these observations suggest promise for tools that explicitly aim to expand people's understandings of decision spaces and framings, as discussed in design goal [\textbf{D5}].

\subsubsection{Is this for me, or the LLM?\nopunct} Participants sometimes expressed uncertainty about whether the system was eliciting and generating information for its own benefit or theirs. In some cases, participants felt that they needed to explain their situation and preferences in as much detail as possible for the system’s sake, to help the system help them. In other cases, participants interpreted the system as having them go through steps mainly to aid their own thinking and reflections. Future research may explore the impacts of such user perceptions in the contexts of AI tools for meta-decision making and ``tools for thought'' more broadly. In either case, participants appreciated how the LLM’s presentation of diverse options suggested that it had not prematurely tried to infer their preferences, instead making it clear that it would take the time to understand them.

\section{Conclusion and Open Questions}

In this short paper, we have outlined an opportunity space for AI-driven tools that support individuals and groups of people in \textit{meta-decision making}. From school admissions to hiring and investment decisions, the development of high-quality criteria and rubrics to guide decision-making is critical. Yet the process of developing decision criteria remains ad hoc and under-supported~\cite{kuo2024policycraft,zhang2023deliberating}.  In this paper, we have presented \textit{InDecision} as a case study of an AI tool to support human meta-decision making. However, we note that \textit{InDecision} represents just one point in a vast design space for AI-augmented meta-decision making. Each of our design goals for AI-augmented meta-decision making points to a set of open questions for future research to explore. As just a few examples: 
\begin{itemize}
    \item Within the broad framework of ``humans as judges and AI as provocateur'' [\textbf{D1}], which steps of iterative criteria development should we be cautious about over-scaffolding? Where is it most important to preserve and protect human thought, unaided by AI? Are there cases where it might be better, for instance, to have humans generate their own self-critiques and provocations, prior to seeing AI-generated provocations? 
    \item   What specific strategies for systematically varying presented examples [\textbf{D2, D3}] and generating potential ``edge cases'' [\textbf{D4}] are most effective in prompting productive reflection that can lead to criteria refinement? Is there additional information that can be seamlessly elicited throughout an iterative criteria development workflow---beyond the sorts of information currently gathered by \textit{InDecision}---that could assist AI tools in generating more targeted provocations?
    \item How can AI tools for meta-decision making best be designed to support \textit{collaboration} when developing decision criteria, in synchronous or asynchronous group settings [\textbf{D6, D7}]? For example, at which steps of an iterative criteria development workflow should groups be encouraged to work together versus separately? And what opportunities are there for tools to leverage differing perspectives and decision patterns between group members to inform AI-generated provocations and facilitate conversations among group members?
\end{itemize}
Overall, we hope that the design goals and initial findings presented in this paper will provide inspiration for future HCI research aimed at helping people develop better strategies for future decisions.

\bibliographystyle{ACM-Reference-Format}
\bibliography{bib}

\appendix

\section{Demo}

For a short video demonstrating \textit{InDecision}, see \url{https://tinyurl.com/indecision-chit4t-demo}.

\end{document}